\documentstyle[aps,epsfig]{revtex}

\begin{document}

\title{Topology, hidden spectra and Bose Einstein-Condensation on low 
dimensional complex networks} 
\author{R. Burioni  \footnote{burioni@fis.unipr.it}, 
D. Cassi \footnote{cassi@fis.unipr.it} and A. Vezzani 
\footnote{vezzani@fis.unipr.it}} 
 
\address{Istituto Nazionale Fisica della Materia (INFM).\\ 
Dipartimento di Fisica, Universit\`a di Parma, parco Area delle Scienze 7A 
43100 Parma Italy} 
 
\maketitle 
 
\begin{abstract} 
Topological inhomogeneity gives rise to spectral anomalies that can induce 
Bose-Einstein Condensation (BEC) in low dimensional systems. These anomalies 
consist in energy regions composed of an infinite number of states with 
vanishing weight in the thermodynamic limit (hidden states). Here we 
present a rigorous result giving the most general conditions for BEC on 
complex networks. We prove that the presence of hidden states in the lowest
region of the spectrum is the necessary and sufficient condition for 
condensation in low dimension (spectral dimension $\bar{d}\leq 2$), while it 
is shown that BEC always occurs for $\bar{d}>2$. 
\end{abstract} 
 
\date{today} 
\pacs{PACS numbers: 89.75-k, 74.80.-g, 5.30.Jp}

The recent experimental evidence for Bose-Einstein condensation 
in real systems \cite{bec} has stimulated an increasing amount 
of theoretical works to find the most general conditions 
inducing such phenomenon in new experimental setups. 
A key problem in this analysis 
is the influence of geometry on the physical behaviour of bosonic models. 
In the classical framework of the ideal Bose gas this influence is encoded 
in the dependence on the space dimension $d$ of the system: 
indeed  BEC occurs if and only if $d\geq 3$\cite{eucl}. 
 
Recently, the interest in quantum devices such as Josephson 
junctions, together with the possibility of combining them in 
complex geometrical arrangements, has stimulated the study of  bosonic 
models on general discrete structures \cite{bh}.
There, the problem of defining an effective dimension describing 
large scale topology has been  successfully solved by the introduction of 
spectral dimension $\bar{d}$ \cite{ao,univ}, which can be experimentally 
measured \cite{misur} and rigorously defined by graph theory \cite{hhw}. 
On the other hand, one expects the influence of topology to be richer and 
more complex on discrete structures, due to the possible relevance of local 
geometrical details, in addition to the large scale structure described by 
dimensionality. 
 
In this  direction, recent works on BEC on inhomogeneous 
networks \cite{beccomb} have put 
into evidence that strong inhomogeneities 
can give rise to condensation at finite temperature even in a 
low-dimensional system as a comb lattice ($\bar{d}=1$, see figure 1). 
This phenomenon arises from a peculiarity of the spectra of inhomogeneous 
networks we shall refer to as {\it hidden spectrum}, 
consisting  of an energy region  filled by a finite of  infinite number of 
eigenvalues which do not 
contribute to the normalized spectral density in the thermodynamic limit. 
Hidden spectra don't usually affect bulk thermodynamic quantities but, 
as we shall show in the following, can have dramatic effect when bosonic 
statistics forces the macroscopic occupation of a single quantum state. 

Here we give a general mathematical definition of hidden spectra and we 
prove that the necessary and sufficient condition for 
condensation when $\bar{d}\leq 2$ is the presence of hidden states in the 
lowest region of the energy spectrum, while BEC at finite temperature 
always  occurs 
for higher dimensional systems (i.e. when $\bar{d}>2$).

A generic discrete network is naturally described by a graph which is  a
countable set $V$ of vertices (sites) $i$ connected
pairwise by a set $E$ of unoriented edges (links) $(i,j)=(j,i)$. We will call
nearest neighbours two vertices joined by an edge.  The graph
topology is algebraically described by its adjacency $A_{ij}=1$ if $(i,j)$ is
a link of the graph and $A_{ij}=0$ otherwise. The coordination number
$z_i=\sum_j A_{ij}$ is the number of nearest neighbours of the site $i$.
A real geometrical structure always has a finite maximum number of nearest
neighbours at any site. To take into account this condition in the
thermodynamic limit we introduce the
uniform boundedness condition: ${\rm max}_i z_i <\infty$ on the
coordination numbers.
A path in  ${G}$ is a sequence of consecutive links
$\{(i,k),(k,h),...,(n,m),(m,k)\}$ and a graph is
said to be connected if for any two point there is always a
path joining them.
In the following we will consider only connected graphs; disconnected graphs
can be reduced to non-interacting connected components which can be studied
separately. Every connected graph is endowed with an intrinsic metric generated
by the chemical distance $r_{i,j}$, which is defined as the number of links in
the shortest path connecting sites $i$ and $j$.

The most general Hamiltonian for non interacting particles
on ${G}$ is:
\begin{equation}
{{H}}=\sum_{i,j\in V} h_{ij}a^{\dag}_ia_j
\label{ham}
\end{equation}
where $a^{\dag}_i$ and $a_i$ are the creation and annihilation operator
at site $i$. The bosonic nature of the particles is introduced
through the usual commutation relations $[a_i,a^{\dag}_j]=\delta_{ij}$.
The Hamiltonian matrix $h_{ij}$ is defined by:
\begin{equation}
h_{ij}= t_{ij} + \delta_{ij} V_i
\end{equation}
The term $t_{ij}$ describes hopping between nearest neighbours sites and
it is directly related to the topology of the graph. Indeed
$t_{ij} \ne 0$ if and only if $(i,j)\in E$ i.e. $A_{ij}=1$. The diagonal term
$V_i$ takes into account a potential at site $i$ and both terms must satisfy a
a boundedness condition: $0<k<|t_{ij}|<K $ and  $0<c<|V_i|< C$.

To introduce the thermodynamic limit and analyze the conditions
for Bose-Einstein condensation on the graph, we begin by studying
models restricted to the Van Hove sphere $S_{r,o}$ of center $o$
and radius $r$. This is defined as the set of the vertices whose distance
from $o$ is equal or  smaller than $r$. We will call $N_{r,o}$ the number of
site in $S_{o,r}$.
The Hamiltonian restricted to the sphere is:
\begin{equation}
{{H}}^{S_{r,o}}=\sum_{i,j\in V} h^{S_{r,o}}_{ij}a^{\dag}_ia_j
\label{hamsr}
\end{equation}
where $h^{S_{r,o}}_{ij}=h_{ij}$ if $i$ and $j$ belong to the sphere and
$h^{S_{r,o}}_{ij}=0$ otherwise. For graphs with
polynomial growth, i.e. when $N_{r,o}\sim r^p$ for $r\to \infty$,
it is possible to show
that the thermodynamic limit is independent from the choice of the center of
the sphere $o$ \cite{rimtim}. In the following we will consider only graphs with
polynomial growth, since this is the necessary condition to ensure that
the structure can be embedded in a finite dimensional space. We will then write:
${}{H}^{r}={}{H}^{S_{r,o}}$, $h^{r}_{ij}=h^{S_{r,o}}_{ij}$ and
$N_{r}=N_{r,o}$.

For each finite sphere of radius $r$ the matrix $h^{r}_{ij}$ defines a
normalized density of states $\rho^r(E)$ which is the sum of $N_r$
$\delta$-functions $\delta(E-E^r_k)$, where $E^r_k$ are the eigenvalues of
$h^r_{ij}$. The density $\rho^r(E)$ will be normalized to $1/N_r$.

In the thermodynamic limit, we define $\rho(E)$ to be
the spectral density of the eigenvalues of $h_{ij}$ if:
\begin{equation}
\lim_{r\to\infty} \int |\rho^r(E)-\rho(E)| dE =0
\label{dens}
\end{equation}
Let us define $E_m \equiv {\rm Inf} ({\rm Supp}(\rho(E)))$ where
${\rm Supp}(\rho(E))$ is the support of the distribution $\rho(E)$. The
asymptotic behavior of the thermodynamic spectral density in this region is
described by the spectral dimension $\bar{d}$ \cite{univ}:
\begin{equation}
\rho(E) \sim (E-E_m)^{{\bar{d}\over 2}-1} \ \ \ \ \ \ {\rm for} \ E\to E_m
\label{defd}
\end{equation}

A {\it hidden} region of the spectrum is an energy
interval $[E_1,E_2]$ such that
$[E_1,E_2]\cap {\rm Supp}(\rho(E))= \emptyset$ and
$\lim_{r\to \infty} N^r_{[E_1,E_2]}>0$, where  $N^r_{[E_1,E_2]}$ is the number
of eigenvalues of $h^r_{ij}$ in the interval $[E_1,E_2]$. Notice that in general
$N^r_{[E_1,E_2]}$ can diverge for $r\to \infty$ and the eigenvalues can become
dense in $[E_1,E_2]$ in the thermodynamic limit. Therefore this condition not
only includes the trivial case of discrete spectrum but is far more general; an
interesting example of this new kind of behaviour has been observed in the comb
lattice \cite{beccomb} (figure 1), where the hidden region of the
spectrum is filled by an infinite number of states.

We now define the lowest energy level for the sequence of
densities $\rho_r(E)$, setting
$E_0^r={\rm Inf}_k (E_k^r)$ and
$E_0=\lim_{r \to \infty} E_0^r$. In general, $E_0\leq E_m$.
If $E_0<E_m$, then $[E_0,E_m]$ is a hidden region of the spectrum,
which will be called hidden low energy spectrum.

In the following we will consider models at fixed fillings
$f=N/N_r$, where $N$ represents the number of particles in the system.
In the macro-canonical ensemble the equation that determines the fugacity $z$ in
the thermodynamic limit is:
\begin{equation}
f= \lim_{r\to\infty}
\int {\rho^r(E) dE\over z^{-1}e^{\beta E}-1}
\label{eq1}
\end{equation}
Setting $E_0=0$ we have that $0\leq z \leq 1$.

The integral in the equation (\ref{eq1}) can be divided into two sums, the
first ones considering the energies smaller than an arbitrary constant
$\epsilon$ and the second the energies larger than $\epsilon$:
\begin{equation}
\int {\rho^r(E) dE\over z^{-1}e^{\beta E}-1}=
\sum_{k=0}^{E_k\leq \epsilon}
{1 \over z^{-1}e^{\beta E_k^r}-1}+
\int _{E>\epsilon} {\rho^r(E) dE\over z^{-1}e^{\beta E}-1}
\label{eq2}
\end{equation}

We define:
\begin{equation}
n_{\epsilon}^r \equiv
\sum_{k=0}^{E_k\leq \epsilon}
{1 \over z^{-1}e^{\beta E_k^r}-1}
\label{ne}
\end{equation}
as the fraction of particles with energy smaller than
$\epsilon$. Bose-Einstein condensation occurs in this systems if it exists a
critical temperature $T_c>0$ such that, for any $T<T_c$,
$n_{\epsilon} \equiv \lim_{r\to\infty}n_{\epsilon}^r>k>0$, for all
$\epsilon>0$; i.e. $n_0\equiv\lim_{\epsilon\to 0}n_{\epsilon}=k>0$.

From the definition (\ref{ne}), $n_0$ can be strictly positive
only if $\lim_{r\to\infty}z(r)=1$. Indeed, if this limit is
smaller than $1$ it follows that:
$n_{\epsilon}\leq(1-z)^{-1}\lim_{r\to\infty} N_{\epsilon}^r / N^r$
where $N_{\epsilon}^r$ is the number of state with energy smaller than
$\epsilon$. If the ground state is not infinitely degenerate,
$\lim_{\epsilon \to 0}\lim_{r\to \infty}(N_{\epsilon}^r/N^r)=0$
and then $n_0=0$.

Taking first the limit $r\to\infty$ and then
$\epsilon \to 0$ in equation (\ref{eq2}) we obtain:

\begin{equation}
f   =   n_0+ \lim_{\epsilon\to 0}  \lim_{r\to\infty}
\int_{E>\epsilon} {(\rho^r(E)-\rho(E))dE\over z^{-1}e^{\beta E}-1} +
\lim_{\epsilon \to 0} \lim_{r\to\infty} \int_{E>\epsilon} {\rho(E) dE\over
z^{-1}e^{\beta E}-1}
\label{eq4}
\end{equation}

Now, from the boundedness of $(z^{-1}e^{\beta E}-1)^{-1}$ for $E>\epsilon$
and from
the definition (\ref{dens}), the first of the two limits in the right hand side
of (\ref{eq4}) vanishes:
\begin{equation}
f= n_0+ \lim_{\epsilon \to 0}
\int_{E>\epsilon} {\rho(E) dE\over z^{-1}e^{\beta E}-1}=
n_0+\int {\rho(E) dE\over z^{-1}e^{\beta E}-1}
\label{eq7}
\end{equation}
where, again, $n_0$ can be different from $0$ only if $z=1$.

Now the integral in equation (\ref{eq7}) is an increasing continuous
function of $z$ with $0\leq z < 1$. If the limit:
\begin{equation}
f_c(\beta)=
\lim_{z \to 1}
\int{\rho(E) dE\over z^{-1}e^{\beta E}-1}
\label{eq8}
\end{equation}
is equal to $\infty$ we have $z<1$ and $n_0=0$. If the limit is
finite, $f_c(\beta)$ is a decreasing function of $\beta$ with
$\lim_{\beta \to \infty}f_c(\beta)=0$ and
$\lim_{\beta \to 0}f_c(\beta)=\infty$.
Then, for a suitable $\beta_C$, $f_c(\beta_C)=f$. For $\beta>\beta_C$,
(i.e. $T<T_C$) $z=1$ and  $n_0=f-f_c(\beta)>0$ while for $\beta<\beta_C$,
(i.e. $T>T_C$) $z<1$ and $n_0=0$.

From the divergence or finiteness of the limit (\ref{eq8}) one obtains
the most general conditions
for the occurrence of Bose-Einstein condensation.

First, if $0=E_0<E_m$ (i.e. the system present a low energy hidden spectrum)
the limit (\ref{eq8}) is finite
and there is Bose-Einstein condensation at finite temperature:

\begin{equation}
f_c(\beta)\leq
{1\over e^{\beta E_m}-1}
\int{\rho(E) dE}={1\over e^{\beta E_m}-1}
\label{eq9}
\end{equation}

On the other hand, when $E_0=E_m$ the value of the limit (\ref{eq8})
is determined by the spectral dimension $\bar{d}$. Indeed if $\bar{d}>2$:
\begin{equation}
f_c(\beta)  \leq
\int_0^{\delta}{c_1 E^{{\bar{d}\over 2}-1} dE\over \beta E}+
\int_{E>\delta} {\rho(E) dE\over e^{\beta E}-1} < \infty
\label{eq10}
\end{equation}
where $\delta$ and $c_1$ are suitable constants. Therefore in
this case Bose-Einstein condensation occurs at finite temperature.

For $d<2$ we have:
\begin{equation}
f_c(\beta) \geq  \lim_{z \to 1}
\int_0^{\delta} { z c_1 E^{{\bar{d}\over 2} -1 } dE\over \beta E +1 -z}=\infty
\label{eq11}
\end{equation}
and there is no Bose-Einstein condensation.
When $\bar{d}=2$ we have to consider the logarithmic correction to (\ref{defd})
and it is possible to show that the limit (\ref{eq8}) diverges.

This result applies to many different situations. The simplest example is the
discretization of the usual Schr\"odinger equation for free particles on
a graph. In this case the Hamiltonian is given by:
$ h_{ij}={\hbar^2 \over 2m} L_{ij}$. $L_{ij}$ is the Laplacian operator
on the graph $L_{ij}=z_i \delta_{ij}-A_{ij}$. It can be shown that $E_0=E_m$
and therefore the occurrence of BEC is determined by the behavior of the
spectrum of $L_{ij}$ at low eigenvalues, described by the spectral dimension
as in (\ref{defd}). $\bar{d}$ is known for a wide class of structures
\cite{hhw,rammal,bundle}: for lattices it coincides with the usual
Euclidean dimension and for exactly decimable graph (the Sierpinski gasket
\cite{rammal} and the T-fractal are illustrated in figure 2) it can be proven
\cite{hhw} that $\bar{d}<2$.

A more important model, relevant for real condensed matter structures,
is a pure hopping of non interacting bosons on graphs. This has been
considered in \cite{beccomb} for the description of the
Josephson junction arrays in the weak coupling limit. In this case the
Hamiltonian matrix is given by: $h_{ij}=-t A_{ij}$.
A relevant point about the application of our result to real systems
concerns the effects of the introduction of a fluctuating local potential.
Indeed it can be shown \cite{prep} that in presence of a hidden spectrum
giving rise to BEC, the existence of a condensate is not affected
by the introduction of a small enough potential.

Let us now focus on the pure hopping model described by the Hamiltonian $h_{ij}=-t A_{ij}$
where the only effects are due to topology, to
show how the general theorem can be applied to a wide class of discrete
structures. Condensation at finite temperature due to the presence of low
energy hidden states is typical of bundle structures \cite{bundle}. These
graphs are obtained by a ``fibering'' procedure, i.e. attaching the origin
site of copy of a graph we will call the ``fiber'' graph, to every point
of another graph called ``base''. An example is the
brush graph, shown in figure 3, where the base (the 2d lattice) is fibered by a
linear chain. On bundled structures the spectrum $\rho(E)$ is simply given by the
spectrum of the pure hopping model on the fiber but the inhomogeneities due to
the base give rise to a low energy hidden spectrum. Moreover, the wave function
of the condensate is localized along the base and presents a fast decay along
the fibers. This result can be in general obtained by first diagonalizing the
Hamiltonian of a pure hopping model defined on the base and then solving
the eigenvalues problem of the
Hamiltonian of the fiber with a suitable impurity in the origin, whose form
depends on the result of the previous diagonalization. (see
\cite{beccomb} for a detailed application of this techniques to the comb graph).
In the case of the brush graph $E_m=\sqrt(20)-2\approx 2.47>E_0=0$ (in all
figures the energy is measured in $t$ units and the energy zero has been
chosen so that $E_0=0$), so that we can apply the general result for low energy
hidden spectra. In the figure the continuous line is $\rho(E)$ and it coincides
with the density of states of the hopping model on a linear chain (the fiber).
The dotted line represents the hidden spectral region, which is filled
continuously by the hidden states. Since on the comb graph the fraction of
states in the hidden spectral region goes to zero as $1/N_r^{1/3}$ for
$r\to\infty$, in order to obtain the dotted lines we have to normalize
$\rho^r(E)$ dividing by $N_r^{2/3}$ and not by the total number of sites
in the sphere.

Graphs with constant coordination number $z_i$ are typical
examples in which the pure hopping model do not present hidden low energy
regions ($E_0=E_m$). This is due to the fact that for this class of graphs
the spectrum of the model can
be obtained from that of the Laplacian matrix by a shift of the zero in
the energy.
The existence of BEC is therefore determined by the spectral dimension
$\bar{d}$ of the graph. This
parameter can be exactly calculated for a wide class of discrete structures. On
lattices, $\bar{d}=d$ is the usual Euclidean dimension and one
recovers the classical result for BEC on translation invariant structures. The
d-dimensional Sierpinski gaskets \cite{rammal} (see figure 2) are example of
exactly decimable graphs with constant coordination number. Since for these
graphs $\bar{d}<2$, our general result proves that in these cases no
Bose-Einstein condensation occurs.

A fundamental property of the low eigenvalues spectral density of the Laplacian
matrix is its independence from the local details of the graph, i.e.
gemoetrical universality \cite{univ}. The value of $\bar{d}$ is
not changed under a wide class of transformations, called isospectralities,
which can strongly modify the geometry of the graph. A simple
consequence of this property is that if we consider the pure hopping model on a
graph with constant coordination number, which differs from a graph of known
dimension $\bar{d}$ by an isospectrality, BEC occurs only if
$\bar{d}>2$. An example of this behaviour is given by the ladder
graph (figure 4) which can be obtained from the linear chain by the addition of
finite-range links, which is one of the simplest cases of isospectralities
On this structure then there is no low energy hidden
spectrum, $\bar{d}=1$ and the pure hopping model does not exhibit BEC.

An important property of $\bar{d}$ is that the dimension of the
graph obtained as a direct product of two graphs is the sum of the
dimensions of the original structures. An
example is illustrated in figure 5,
where we show the direct product of a linear chain and a well known
fractal, a Sierpinski gasket. In this case the coordination number
$z_i$ is constant. Applying this property of $bar{d}$ it is easy to show
that $\bar{d}=1+2\ln(3)/\ln(4)>2$, and therefore  one immediately
infers that BEC occurs on this graph. Interestingly, in this case we do not know the
analytical form of the
spectrum of the Hamiltonian (the spectral density in the figure is obtained by a
numerical diagonalization), nevertheless we can prove the existence of BEC from
the general theorem and from the properties of $\bar{d}$.

\begin{figure}[h]
\centerline{\psfig{
figure=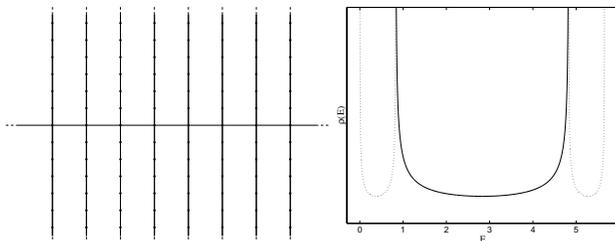,width=82mm,angle=0}}
\caption{The comb graph and the spectrum of the pure hopping model 
($h_{ij}=-tA_{ij}$) defined on this structure. The hidden spectrum is 
represented with dotted lines. In this case the spectrum is obtained from 
an exact calculation in the thermodynamic limit. 
The energy is measured in units of $t$ and the zero has 
been chosen so that $E_0=0$; the density is plotted in arbitrary units 
(the hidden spectrum and $\rho(E)$ can be normalized to 1 dividing 
respectively to the number of states belonging to each spectral region).}
\end{figure}

\begin{figure}[h]
\centerline{\psfig{
figure=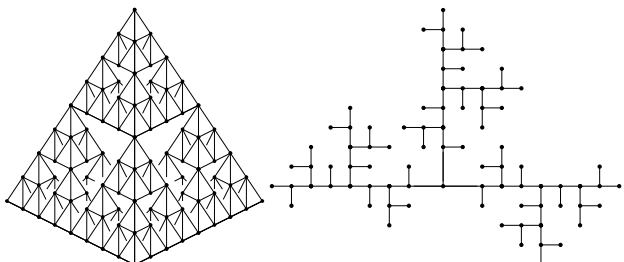,width=82mm,angle=0}}
\caption{Two classical examples of exactly decimable graph: the 3d Sierpinski 
gasket ($\bar{d}=2\ln(4)/\ln(6)<2$) and the t-fractal 
($\bar{d}=2\ln(3)/\ln(6)<2$).}
\end{figure}

\begin{figure}[h]
\centerline{\psfig{
figure=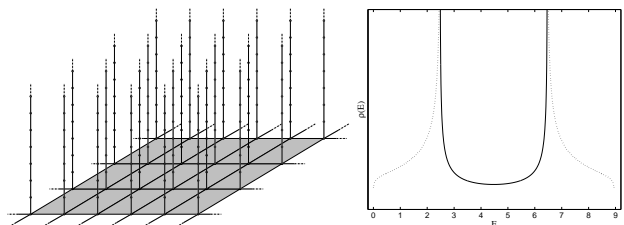,width=82mm,angle=0}}
\caption{The brush graph and the spectrum of the pure hopping model 
on this graph (obtained from the diagonalization of a structure of 
$200\times 200 \times 200$ sites). The dotted lines represent the 
hidden spectrum which gives rise to condensation even if the spectral 
dimension is 1.}
\end{figure}

\begin{figure}[h]
\centerline{\psfig{
figure=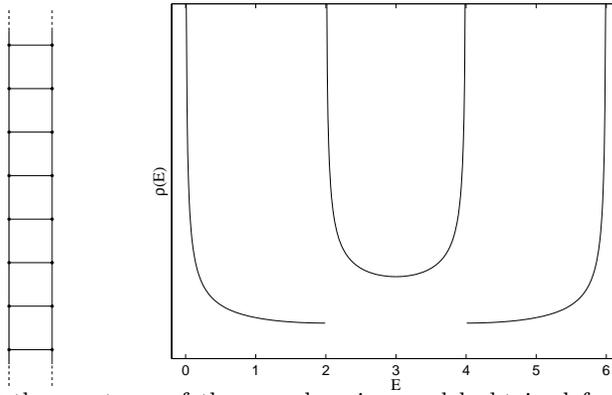,width=82mm,angle=0}}
\caption{The ladder graph and the spectrum of the pure hopping model 
obtained from an exact diagonalization in the thermodynamic limit. 
Here there is not hidden spectrum and $\rho(E)\to \infty$ when 
$E\to E_m=E_0$ ($\bar{d}=1$), then there is not condensation at finite 
temperature.}
\end{figure}

\begin{figure}[h]
\centerline{\psfig{
figure=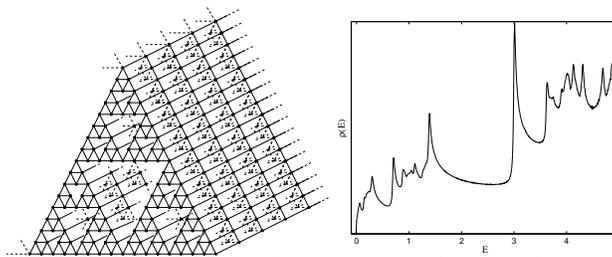,width=82mm,angle=0}}
\caption{The graph obtained as a product of the Sierpinski gasket and 
a linear chain. Here the spectrum of the pure hopping model is obtained from 
the diagonalization of a graph of 840192 sites. Here condensation occurs 
since $\bar{d}=1+2 \ln(3)/\ln(4)>2$ (indeed the spectral vanishes for 
$E\to E_m=E_0$).}
\end{figure}

\end{document}